# Disaster-Resilient Control Plane Design and Mapping in Software-Defined Networks

S. Sedef Savas[1], Massimo Tornatore[1, 2], M. Farhan Habib[1], Pulak Chowdhury[1], and Biswanath Mukherjee[1]

[1]University of California, Davis, USA  [2]Politecnico di Milano, Italy

*Abstract*—Communication networks, such as optical core networks heavily depend on their physical infrastructure, and hence they are vulnerable to man-made disasters, such as Electromagnetic Pulse (EMP) or Weapons of Mass Destruction (WMD) attacks, as well as to natural disasters, such as earthquakes. Large-scale disasters may cause huge data loss and connectivity disruption in these networks. As society's dependence on network services increases, the need for novel survivability methods to mitigate the effects of disasters on communication networks becomes a major concern. Software-Defined Networking (SDN), by centralizing control logic and separating it from physical equipment, facilitates network programmability and opens up new ways to design disaster-resilient networks. On the other hand, to fully exploit the potential of SDN, along with data-plane survivability we also need to design the control plane to be resilient enough to survive network failures caused by disasters. For resiliency of the control-plane, we need to select appropriate mapping of the controllers over the physical network, and then ensure that the connectivity among the controllers (controller-to controller) and between the controllers and the switches (switch to controllers) is not compromised by physical infrastructure failures. Several distributed SDN controller architectures have been proposed to mitigate the risks of overload and failure, but they are optimized for limited faults without addressing the extent of large-scale disaster failures. In this paper, we present a novel disaster-aware control-plane design and mapping scheme, formally model this problem, and demonstrate a significant reduction in the disruption of controller-to-controller and switch-to-controller communication channels using our approach.

## I. INTRODUCTION

Disasters events (e.g., due to Electromagnetic Pulse (EMP) and Weapons of Mass Destruction (WMD) attacks, earthquakes, hurricanes, etc.) represent a challenging threat for communication networks, such as optical backbone networks, as they affect large geographical areas and may cause multiple network failures in the disaster zone. These failures could also be cascading, i.e., when a disaster occurs, initially a set of network elements may fail simultaneously, and then other failures in different parts of the network may occur subsequently (e.g., due to a power outage following an earthquake). Especially, targeted events, such as EMP attacks [1] and WMD attacks, tend to be more catastrophic compared to a random natural disaster.

Recent disaster events have shown the enormous loss of network resources caused by both initial failures and correlated cascading failures [2]. For instance, in the 2008 Shichuan earthquake, around 30,000 km of fiber optic cables and 4,000 telecom offices were damaged [3]. In the 2011 Japan Earthquake and Tsunami, around 1,500 telecom buildings experienced long power outages by the main shock on March 11; while most were fixed, 700 telecom buildings experienced power outages by the aftershock on April 7, 2011 [2]. Because of the potential huge data loss and network-connectivity disruptions, high-capacity optical communication networks must be designed to provide resiliency against disasters, be adaptable to rapidly-changing network conditions due to disasters, and be able to recover after disasters, even if they are infrequent.

Today's networks are inflexible to effectively respond to large, complex disaster failures. They are rigidly vertically-integrated, i.e., control and data planes are bundled together, and this is a main reason behind the complexity of network reconfiguration and management. Software-defined networking (SDN), a new networking paradigm, has emerged as a promising candidate to support the new demands and requirements of current and future communication networks. SDN breaks vertical integration of traditional networks by separating the network's control logic from the underlying routers and switches, promotes (logical) centralization of network control, and introduces the ability to program the network. The evolution of traditional control logic towards a centralized control plane simplifies network management, facilitates new vendor-independent network innovations, and introduces new optimization opportunities with the global view it provides.

These new functions of SDN can be exploited to provide higher survivability against disasters, but these opportunities come with a cost. In SDN, along with data-plane survivability, we also need to design the control plane and its communication with the data plane to be resilient enough to survive disaster disruptions.

Early deployments of SDN relied on physically-centralized control-plane architectures with a single controller, but a centralized system suffers from scalability, performance, and reliability problems in a large network deployment. Even a very powerful controller will lack the CPU and memory capacity necessary to maintain complete network state, and react to all network events, for large, high-capacity networks [5]. Also, the centralized design is vulnerable to disruptions and attacks, particularly to single point of failures [6]. Several distributed SDN controller architectures have been proposed to mitigate the risks of overload and failures. However, disaster-resilient design of control plane is almost unexplored, and it poses several challenges. As disasters occur in specific geographical locations and disrupt specific parts of the network, design of the network determines the impact of the disasters. Ignoring the vulnerable regions of the physical network while designing the distributed control plane and assigning switches to controller instances increases the risk of disconnection between switches and controllers, as well as controller-to-controller communication. Reprovisioning these communication channels can cause huge time loss in the event of disasters, and eventually connection disruption rate increases. To address this, we present a new approach to design the control plane in a disaster-aware manner by considering it as a virtual network, and solving it using a modified virtual-network-embedding approach. Our scheme decides on the number and placement of the controllers, maps them onto the physical network, and assigns switches to distributed controller instances in a disaster-resilient way.

The rest of this study is organized as follows: Section II reviews previous works on disaster resilience in communication networks and SDN. Section III presents the proposed disaster-resilient control-plane design framework, and its mathematical formulation is provided in Section IV. Section V describes the simulation scenario and presents the obtained results. Finally, Section VI concludes the study.

## II. RELATED WORK

Researchers have proposed a number of techniques to overcome the disadvantages of the physical centralization of the SDN control plane in terms of not being robust against overloads and failures. Proposals exist to push the intelligence into the switches to offload the controller [7][8] and to offer a basic connectivity in the data plane [9]. Also, a logically-centralized yet physically-distributed control plane as in Fig. 1 has been built such as Kandoo, HyperFlow, and Onix [10], which can benefit from the scalability and reliability of a distributed architecture while preserving the simplicity of a centralized system. Besides providing higher survivability against network failures, advantages of the distributed control plane include more scalability (control-plane throughput increases as many controllers can be used for load balancing) and reduced control delay by choosing the quickest-responding controller.

The problem of how many controllers to use and where to place them in the distributed control-plane design was studied in [11] using a static setting and with a fixed number and placement of controllers. Refs. [10][12] propose dynamic controller provisioning where number of controllers and their locations change dynamically with changing network conditions to minimize flow setup time and communication overhead.

Nonetheless, even considering the inherent survivability of a distributed control-plane system, controller distribution does not guarantee control-plane network connectivity, which is a necessity for SDN to function properly against network failures, especially after a large-scale disaster failure. Existing distributed control-plane designs do not address the extent of large-scale disasters, despite the catastrophic disaster effects on communication networks that have been experienced in the past.

The first step in disaster survivability is *modeling the disaster*. Several studies (e.g., [13]) aim at modelling disasters and define the parts of the network that are more vulnerable to regional/correlated failures caused by disasters for analysis and/or design purposes. Although there are many studies that exploit the information on vulnerable regions of the network to proactively (before disasters occur) and/or reactively (after disasters) take necessary actions to minimize the loss in *data plane*, disaster resilience in SDN *control plane* is largely unexplored.

In this work, we study the problem of designing a disaster-resilient SDN-based control plane to make the control plane resilient and dynamically adaptable to all disruptions: controller failure, inter-controller communication failure, and controller-to-switch communication.

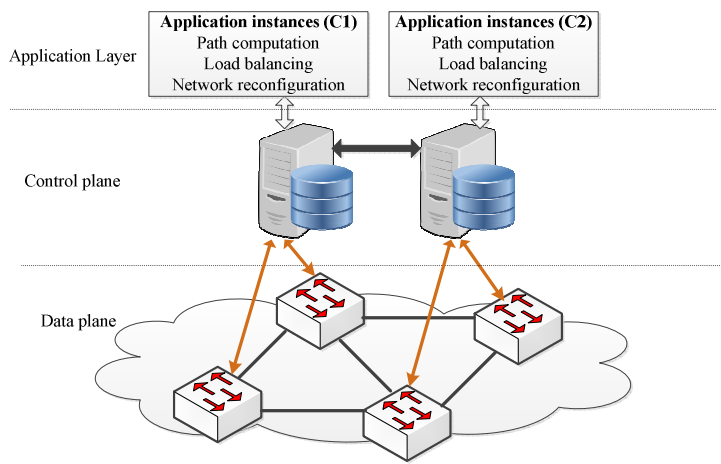

Fig. 1. Distributed control-plane architecture.

## III. DISASTER-AWARE CONTROL-PLANE DESIGN

A distributed control-plane topology can be designed as an overlay (i.e., virtual/logical) network mapped over a physical (i.e., backbone) network, which we call the *Control Network Mapping* (CNM) problem. This mapping problem can be solved using *virtual network embedding* (VNE) [14-15] techniques which are defined as the assignment of virtual network resources to physical network elements, e.g., virtual links are created using multi-hop physical links. CNM allocates necessary resources in the substrate network through node mapping and link mapping. Different from the traditional VNE problem, in CNM, a virtual network (control network in this case) topology is not given, i.e., the selection of the number of virtual nodes (controllers) and of the virtual links connecting them are also a part of this new problem. The selection of virtual links and nodes, and their mappings are jointly optimized, because preselecting a virtual topology without considering its relation to the link-mapping phase restricts the solution space and can result in poor performance. Given these additional decisions, existing schemes already developed for Wavelength-Division Multiplexing (WDM), Orthogonal Frequency Division Multiplexing (OFDM) or Layer 2/3 VNE solutions are not directly applicable to our problem.

In Fig. 2(a), a control plane is mapped onto a physical topology while minimizing resource usage, whereas in Fig.

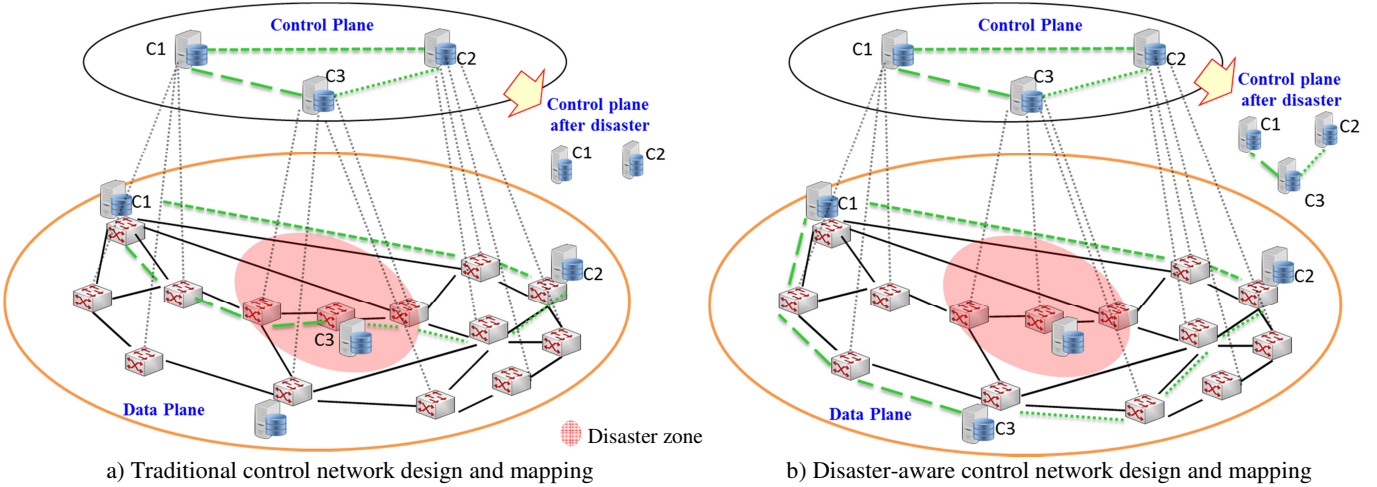

a) Traditional control network design and mapping  b) Disaster-aware control network design and mapping

Fig. 2. Control network mapping (CNM).

2(b), the control plane is mapped considering a predicted disaster zone. After a disaster occurs, in the disaster-unaware design, one of the controllers (C3) fails, and hence the switches connected to it get disconnected from the control logic of the network. Also, the control plane gets islanded, and the switches connected to the controllers located on different islands can no longer communicate. Although the physical topology is still connected, due to the control-plane disconnection, the data plane also gets islanded.

### 1. Disaster Risk Modeling

To devise a disaster-resilient mapping the control plane, network vulnerability to different types of disasters in different locations needs to be assessed. In this study, we use a *probabilistic* disaster model [13], where a network component inside the affected region of a disaster fails with a probability that depends on many factors such as its distance from the disaster's epicenter, and type and magnitude of the disaster. Using this model, we calculate a vulnerability metric, called *disaster risk*, to be used to design a disaster-resilient control plane. *Disaster risk* captures the expected connectivity loss of the control plane due to controller node failure, inter-controller communication failure, and controller-to-switch communication failure given the location, span, and probability of occurrence of the potential disasters. *Risk* is calculated by the formula:

$$\sum_{y \in Y} \sum_{p \in P} R_p^y \cdot p_y^{fail} \qquad (1)$$

where $Y$ is the set of all disasters; $P$ is set of all paths used for controller-to-controller and controller-to-switch communications; $R_p^y \in \{0,1\}$: 1 if path $p$ is affected in case of disaster $y$; and $p_y^{fail}$ is the probability of occurrence of disaster $y$ multiplied by probability that disaster $y$ causes a failure. This formula covers controller node failures because, when a node is down, paths passing through that node also get affected.

### 2. Problem Statement

We formalize and investigate this new problem of CNM to ensure control-plane connectivity against both single point of failures and large-scale disaster failures. We present a mathematical formulation for jointly optimizing the virtual topology design and virtual network embedding such that control-network connectivity is ensured after failures.

This problem can be stated as follows: *Given the network topology with the nodes (switches) connected to a datacenter (which are possible locations for controller placement) and survivability requirements of the network, find the number of controllers, decide on the virtual topology, and perform virtual control network mapping such that the network is survivable against any single point of failure and predicted disaster failures while minimizing physical resource usage*. While minimizing the resource usage, we also consider inter-controller and switch-to-controller delay to be able to respond to failures promptly. The ultimate aim is to make the control plane resilient to controller failure, inter-controller communication failure, and controller-to-switch communication failure.

## IV. PROBLEM FORMULATION

We formulate the problem of designing a control plane (deciding on the number of controllers, their placement, and the control-plane topology) and mapping it on a physical network as an integer linear program (ILP), given below. To simplify the model, we assume that physical links have no constraints on bandwidth resource. Also, we assume control communication is in-band (i.e., control communication channels use data-plane resources). A subset of the switches has datacenters located onsite, where controllers can be deployed. We evaluate the *disaster risk* of paths between switch to controller and controller to controller offline. Based on the risk information, our formulation decides the number and placement of controllers, maps the control plane, and performs controller-to-switch assignments.

We ensure control-plane connectivity after the predicted disaster occurrence by assigning two disaster-zone-disjoint virtual paths for each controller-to-controller communication.

We formulate the problem as follows:
*Given:*
- $G(N,E)$: Network topology where $N$ is the set of nodes and $E$ is the set of directed links.
- $F$: Set of network nodes which are connected to a datacenter where a controller can be deployed. Only these switches can host a controller.
- $d_{ij} \in \{0,1\}$: 1 if node $j$ is located within node $i$'s reachability island. Reachability island of a node is a

circular region where shortest-path distance to every node within this region from this node satisfies the latency constraint of this node. Latency constraint can be determined by the network operator and depends on the required time to react a particular event (such as connection arrival, departure, etc.)

- $k$: Number of controllers that must be guaranteed to be located within a certain latency limit by every switch.
- $q$: At least $q$ controllers are active at any time in the network.
- $B$: Maximum number of switches that can be assigned to any controller.
- $P_{ij}$: Set of possible paths to be used for virtual-link mapping between node $i$ and node $j$.
- $U_p^y \in \{0,1\}$: 1 if path $p$ survives disaster $y$.
- $Y = \{y \mid y = <E_y, \rho_y>\}$: Set of disasters where $E_y$ is the set of links that are part of Disaster $y$ and $\rho_y$ is the probability that disaster $y$ causes a failure.

*Variables:*
- $C_f \in \{0,1\}$: 1 if a controller is deployed and active on node $f$.
- $V_{st} \in \{0,1\}$: 1 if there is a virtual link between controllers $s$ and $t$.
- $a_{if} \in \{0,1\}$: 1 if switch $i$ is assigned to controller $f$.
- $m_{ij}^{st} \in \{0,1\}$: 1 if virtual link between controller $i$ and controller $j$ carries flow for $(s,t) \in (F,F)$.
- $o_{ifst} \in \{0,1\}$: 1 if nodes $i$, $f$, $s$, and $t$ are controllers.
- $A_{st}^p \in \{0,1\}$: 1 if path $p$ is used for the virtual link between controller $s$ and controller $t$.
- $X_{st}^p \in \{0,1\}$: 1 if path $p$ is used for the virtual link between controller $s$ and switch $t$.
- $K_{if}^y \in \{0,1\}$: 1 if virtual link $V_{if}$ survives disaster $y$.
- $K_{if}^{sty} \in \{0,1\}$: is 1 if virtual link between controller $i$ and controller $f$ carries flow for $(s,t) \in (F,F)$ after disaster $y$

*The objective function* below minimizes the risk of communication channel disruption in the control plane:

Minimize (Min-Risk Optimization)

$$\sum_{y \in Y} \sum_{p \in P} \sum_{i \in N} \sum_{f \in N} (A_{if}^p \cdot U_p^y + X_{if}^p \cdot U_p^y) p_y^{fail}$$

where $p_y^{fail}$ is the probability of occurrence of a disaster.

The resource minimization (Min-Resource) objective function of the disaster-unaware scheme for comparison purposes is as follows:

Minimize (Min-Resource Optimization)

$$\sum_{p \in P} \sum_{i \in N} \sum_{f \in N} (A_{if}^p + X_{if}^p) \cdot Length\_of\_p$$

where the resource usage is calculated as the total number of links used for the virtual link mappings between controller-to-controller and controller-to-switch communication.

*Reachability Constraint:*

$$\sum_{f \in F} C_f \cdot d_{if} \geq k \quad \forall i \in N \qquad (1)$$

It ensures that at least $k$ controllers are active within the reachability island of any switch node (in our case $k=2$).

*Binarization of $C_f$ Variable:*

$$C_f \leq \sum_{i \in N} a_{if} \qquad \forall f \in F \qquad (2a)$$
$$C_f \geq \sum_{i \in N} a_{if} / M \qquad \forall f \in F \qquad (2b)$$

where M is a very large number. All controllers which have an assigned switch should be up and running.

*Controller-Capacity Constraint:*

$$\sum_{i \in N} a_{if} + 1 \leq B \qquad \forall f \in F \qquad (3)$$

Total number of switches assigned to a controller should not exceed a maximum number to avoid overloading the controllers. We assume that each switch which has a controller on it is automatically assigned to that controller.

*Switch-Assignment Constraint:*

$$a_{if} \geq C_f \quad \forall i, f \in N \text{ when } i = f \qquad (4a)$$
$$\sum_{i \in F} a_{if} = 1 - C_f \qquad \forall f \in N \qquad (4b)$$

Every switch is assigned to exactly one controller within latency limits. If there is a controller located at a switch location, that switch is assigned to it, along with the other switches that are assigned to that controller.

*Latency Constraint:*

$$a_{if} \leq d_{if} \quad \forall i \in N, \forall f \in F \qquad (5)$$

A switch only can be assigned to a controller that is located within latency limits. If a controller is not within the reachability distance of a switch, it automatically sets $a_{if}$ to 0.

*Switch to Controller Assignment Check:*

$$a_{if} \leq \neg C_i \wedge C_f \quad \forall i \in N, \forall f \in F \qquad (6)$$

A switch can only be assigned to a node which has a controller on it.

*Setting Virtual Links Between Controller Nodes:*

$$o_{ifst} = C_f \wedge C_i \wedge C_s \wedge C_t \quad \forall f, i, s, t \in F \qquad (7a)$$
$$m_{if}^{st} \leq o_{ifst} \quad \forall f \in F \; \forall i \in F \; \forall (s,t) \in (F,F) \qquad (7b)$$

These constraints ensure that the virtual links are only selected between controller nodes. Note that, Eqns. (6) and (7a) can be linearized, but such linearization is not reported due to space limit.

*Flow-Conservation Constraint:*

$$\sum_{f \in F} m_{if}^{st} - \sum_{f \in F} m_{fi}^{st} \begin{cases} \geq 2, & i = s \\ \leq 2, & i = t \\ = 0, & ow \end{cases} \quad \forall i, s, t \in F \qquad (8)$$

These constraints ensure that there will be at least two disjoint paths between every controller pair to be resilient against a single point of failure (node/link). These are the paths of the virtual control-plane topology.

*Virtual Links Determination Constraint:*

$$V_{if} \leq \sum_{s \in N} \sum_{t \in N} m_{if}^{st} \quad \forall f, i \in F \qquad (9a)$$
$$V_{if} \geq \sum_{s \in N} \sum_{t \in N} m_{if}^{st} / M \quad \forall f, i \in F \qquad (9b)$$

Based on the flows enforced in Eqn. (8), this constraint ensures that the virtual links are set between controllers.

*Controller- to-Controller and Controller-to-Switch Virtual Link Mapping to Physical Topology*

$$\sum_{p \in P_{if}} A_{if}^p = V_{if} \quad \forall i, f \in F \quad (10a)$$

$$\sum_{p \in P_{if}} X_{if}^p = a_{if} \quad \forall i, f \in F \quad (10a)$$

This constraint ensures that one physical path should be assigned for each virtual link between any pair of controllers.

*Constraint to Determine if Virtual Links are Affected by a Disaster:*

$$\sum_{p \in P_{if}} A_{if}^p \cdot U_p^y = K_{if}^y \quad \forall y \in Y \quad \forall i, f \in F \quad (11a)$$

$$K_{if}^{sty} \leq K_{if}^y \quad \forall i, f, s, t \in F \; \forall y \in Y \quad (11b)$$

If the physical path $A_{if}^p$ over which the virtual link *i-f* is mapped survives from the disaster, this means virtual links survive too. This constraint determines the controller-to-controller virtual links survive a disaster. Binarization of $K_{if}^y$ variable is also performed.

*Survivability Constraint:*

$$\sum_{f \in F} K_{if}^{sty} - \sum_{f \in F} K_{if}^{sty} = \begin{cases} +1, & i = s \\ -1, & i = t \\ 0, & ow \end{cases} \quad \forall i, s, t \in F \; \forall y \in Y \quad (19)$$

After any disaster, this constraint ensures that at least one path (composed of virtual links) survives between any controller pair. It is similar to the flow-conservation constraint, but it ensures post-disaster, and not the normal, operation.

ILP is for static problems, and do not scale for large problem instances. By developing heuristic methods, this problem can be adaptable to a dynamic scenario where we can redesign and remap the control network (possibly by restricting the amount of allowable changes in the logical topology) when network state changes (e.g., when intelligence agencies predict an attack, risky regions in the network also change).

## V. ILLUSTRATIVE NUMERICAL EXAMPLES

We study a 14-node NSF physical network (Fig. 3) with 32 Gbps link capacity. As the disaster scenario, we consider a large-scale EMP attack, which affects four nodes and seven links. We chose a drastic example to show that, even when large portion of the network goes down, our scheme can still make the remaining network functional by preserving connectivity for the control communication channels. There are six nodes attached to datacenters, namely nodes 1, 3, 5, 8, 10, and 13, which are possible controller locations. For the latency constraint, each switch is allowed to connect only to controllers located within a 3-hop distance. If a switch has a controller on it, it will be assigned to that controller. We compare our disaster-aware control-plane design with a traditional (disaster-unaware) scheme which designs the control plane and its communication with the switches based on resource consumption minimization while providing protection against single-link failures.

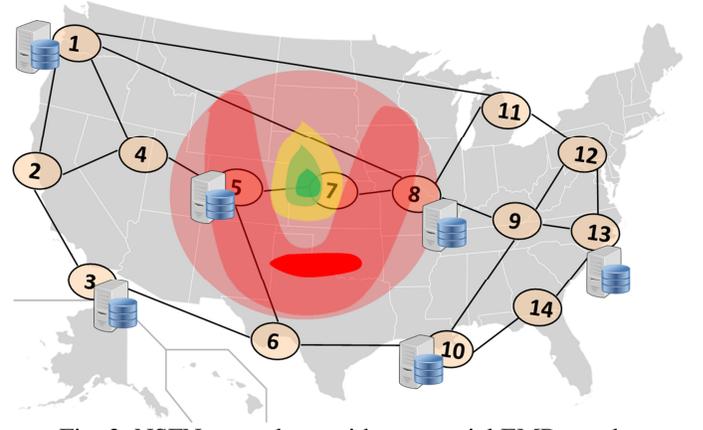

Fig. 3. NSFNet topology with a potential EMP attack. Different colors show EMP fields with different strengths.

In the proposed scheme, risk is minimized using the objective function (Min-Risk) in Section IV. In Fig. 4, we compare these two schemes in terms of number of disruptions experienced in the control plane after a disaster. Figure 4 shows i) the number of virtual links between switches and controllers that fail due to physical network equipment failure, and ii) the number of failed nodes where a controller is located for an increasing number of controllers deployed. We see that, as long as there are datacenters available outside of the disaster zone that meet latency requirements of the network, controllers are placed at these locations; but, in the disaster-unaware scheme, controllers are placed in datacenters located at high-risk nodes.

In Fig. 5, resource consumptions of the two schemes are compared in terms of total number of links that are used for control communication. Our evaluations show that our approach (Min-Risk) yields a significant decrease in the number of disconnected switches with an increase in the capacity required. The increase in capacity requirement is correlated with the predicted disaster scale. Protecting the network from an EMP attack can be costly due to its extremely large scale, and even in this dramatic scenario, our design requires 25% more resources compared to disaster-unaware design (Min-Resource) when three controllers deployed (see Fig. 5), to save as much as 50% of the switches getting disconnected from the control plane.

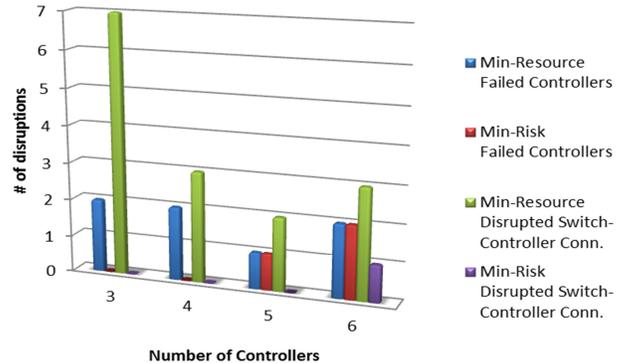

Fig. 4. Comparison of disaster-aware (Min-Risk) and disaster-

unaware (Min-Resource) schemes in terms of disruptions on the control plane caused by an EMP attack.

In this setting, three controllers were enough for disaster-aware design (Min-Risk) to provide the lowest risk possible. When we force the design to have more controllers, some nodes in risky zones are also used as controllers, and this increases disruptions. For the disaster-unaware scheme (Min-Resource), although more controllers do not necessarily mean better survivability, still we observe a sudden decrease in disruptions when we increase the number of controllers. The connections to and from the nodes that are failed during the disasters are not counted as disrupted.

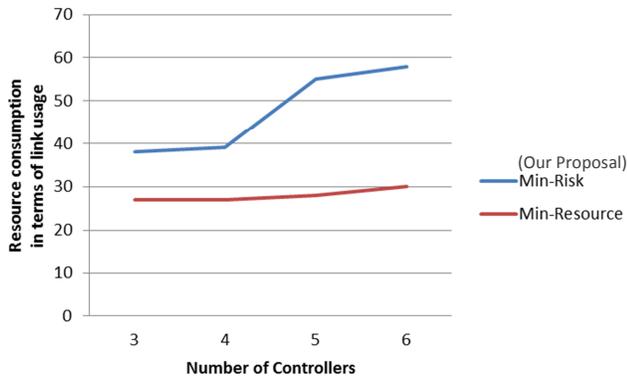

Fig. 5. Comparison of resource consumption of control plane and switch-to-controller communication channels.

## VI. CONCLUSION AND FUTURE WORK

In this paper, we have shown that the current disaster-unaware distributed control-plane architectures are not sufficient to provide resilience in the event of disasters due to disconnections in the control network. We presented a disaster-aware, efficient and distributed SDN control plane that jointly minimizes the control-plane disconnections due to disasters and resource consumption for the control-plane network. Our solution determines the number of controllers, and their placement along with control-plane topology and its mapping to the physical network. Our initial evaluation results are promising, and we plan to develop a practical heuristic that solves the problem within strict time constraints since solving this problem is NP-hard. Also, the adaptability to changing network conditions due to disasters needs investigation.

### ACKNOWLEDGEMENT

This work has been supported by the Defense Threat Reduction Agency (DTRA) Grant No. HDTRA1-14-1-0047.